\documentclass[a4paper,11pt]{article}
\usepackage{a4wide,amsmath,amssymb,bbm}
\usepackage[english]{babel}

\usepackage{cite}

\makeatletter
\g@addto@macro\bfseries{\boldmath}

\@addtoreset{equation}{section}
\makeatother

\begin{document}

\hfill{ 
}

\vspace{30pt}

\begin{center}
{\Huge{\bf Constraining integrable AdS/CFT with factorized scattering}}

\vspace{50pt}

{\bf Linus Wulff}

\vspace{15pt}

{\it\small Department of Theoretical Physics and Astrophysics, Masaryk University, 611 37 Brno, Czech Republic}\\

\vspace{100pt}

{\bf Abstract}
\end{center}
\noindent
We consider (warped) AdS string backgrounds which allow for a GKP spinning string/null cusp solution. Integrability implies that the worldsheet S-matrix should factorize, which in turn constrains the form of the warp factor as a function of the coordinates of the internal space. This constraint is argued to rule out integrability for all supersymmetric $AdS_7$ and $AdS_6$ backgrounds as well as $AdS_5$ Gaiotto-Maldacena backgrounds and a few highly supersymmetric $AdS_4$ and $AdS_3$ backgrounds.

\pagebreak 
\tableofcontents

\setcounter{page}{1}


\section{Introduction}
Integrability is a very powerful tool in the context of the AdS/CFT correspondence \cite{Maldacena:1997re} which allows, at least in principle and at large $N$, for the solution of the theory, see \cite{Beisert:2010jr,Bombardelli:2016rwb} for reviews. It is therefore of great interest to search for new integrable examples. It is in general easier to rule out integrability than to try to prove integrability and this is the strategy we will follow here.

One strategy to rule out integrability of strings in a given background is to find a classical solution of the string equations of motion and show that the equations for the fluctuations around this solution are not integrable. The easiest way to do this is to find a consistent truncation to a one-dimensional problem. This leads to a so-called normal variational equation for which there is a standard procedure to demonstrate (non)integrability. This approach was pioneered in \cite{Zayas:2010fs} and has since been applied to several examples \cite{Basu:2011dg,Basu:2011fw,Basu:2012ae,Stepanchuk:2012xi,Chervonyi:2013eja,Giataganas:2013dha,Giataganas:2014hma,Asano:2015eha,Asano:2015qwa,Panigrahi:2016zny,Basu:2016zkr,Ishii:2016rlk,Giataganas:2017guj,Roychowdhury:2017vdo,Nunez:2018ags,Nunez:2018qcj}. This is often referred to as the analytic integrability approach.

Here we will follow another strategy proposed in \cite{Wulff:2017hzy,Wulff:2017lxh}. This approach is based on the fact that the S-matrix of an integrable massive 1+1 dimensional theory must factorize into a product of two-particle S-matrices \cite{Zamolodchikov:1978xm,Parke:1980ki}. In fact the set of momenta and masses must be conserved in any scattering process. This follows from the fact that an integrable theory has higher conserved charges, which in the free case are just higher powers of the momentum. Since the particles can be assumed to be well separated in the in- and out-state the total conserved charge before(after) is just the sum of the individual free particle conserved charges and this implies that the set of masses and momenta must be the same before and after the collision. The reason that this logic applies only to scattering of massive particles is that only in that case can one ensure that the particles in the in(out) state can always be separated. In a massless theory the existence of higher conserved charges does not in general imply that scattering factorizes, in fact in many theories it does not \cite{Hoare:2018jim}.

Here we will use only the simplest of these conditions, that a process $xx\rightarrow yy$ is forbidden in an integrable theory if the massess of $x$ and $y$ are different.\footnote{A word of caution is in order here. The known proofs that higher conserved charges imply factorization of scattering, e.g. \cite{Parke:1980ki}, assume a Lorentz invariant theory with a strictly massive spectrum. In the string theory applications we have in mind we have neither of these. While Lorentz-invariance is broken only by the interactions and the massless particles don't contribute to the amplitudes we compute, the correspondence between integrability and factorization of scattering is to be viewed as an assumption.} To have a worldsheet S-matrix to talk about we must first pick a classical string solution, or vacuum, around which we expand. Since we are interested in $AdS$ backgrounds a natural choice is the GKP spinning string \cite{Gubser:2002tv}. For technical reasons it is easier to work with its Euclidean version \cite{Kruczenski:2007cy}, the so-called null cusp solution \cite{Kruczenski:2002fb}. It will be enough to consider the bosonic string in light-cone gauge \cite{Metsaev:2000yf,Metsaev:2000yu} which we then expand around the null cusp solution. This gives us a two-dimensional theory where we can perform perturbative calculations \cite{Giombi:2009gd}. In particular we are interested in tree-level two-particle scattering. Similar calculations were performed for the $AdS_5\times S^5$ case in \cite{Bianchi:2015iza}. The difference in the present case is that, to allow for the most general backgrounds relevant to AdS/CFT, we consider warped $AdS$ backgrounds where the metric takes the form
\begin{equation}
ds^2=e^{2A}ds^2_{AdS_n}+ds^2_{M_{10-n}}\,,
\label{eq:warped-metric}
\end{equation}
where the warp factor $A(x)$ is in general a non-trivial function of the coordinates of the compact space $M_{10-n}$. Expanding around the null cusp solution we have\footnote{A linear term in $x$ would not be consistent with the null cusp solution.}
\begin{equation}
e^{2A}=1+2a_{ij}x^ix^j+\mathcal O(x^3)\,.
\label{eq:e2A}
\end{equation}
This leads to mass terms for the $x$ fields, the mass-squared being given by the eigenvalues of the symmetric matrix $a_{ij}$. These masses will in general be different from the masses of the $AdS$ excitations and therefore integrability requires the amplitude for two $AdS$ excitations to annihilate into two internal space excitations to vanish. As we will see this means that the eigenvalues of $a_{ij}$ can only be 0 or $\frac12$ (in suitable units). The only exception is $AdS_3$ where eigenvalue $1$ is also (trivially) allowed by our analysis. The eigenvalue being zero looks, to the order we are working here, like having no warping so it is clear that in this case there cannot be any further restriction. How can we understand that eigenvalue $\frac12$ is allowed? This is easily seen by noting that $AdS_{n+1}$ can be written as a warped product of $AdS_n$ and the line
\begin{equation}
ds^2_{AdS_{n+1}}=\cosh^2{x}\,ds^2_{AdS_n}+dx^2\,,
\end{equation}
and in this case the warp factor has precisely the expansion corresponding to eigenvalue $\frac12$. Clearly also in this case we should not get any extra condition since we are really just in normal unwarped $AdS$. Our findings can therefore be summarized by saying that, provided there exists a GKP solution, integrability requires the warping to be absent, at least locally, up to quadratic order in the coordinates (with the possible exception of $AdS_3$).

We will see that this simple condition on the eigenvalues of $a_{ij}$ in (\ref{eq:e2A}) rules out\footnote{At least in any standard realization and modulo the possible existence of $AdS_7$ and $AdS_6$ solutions where the warp factor does not have a regular critical point.} integrability for supersymmetric $AdS_7$ and $AdS_6$ backgrounds, $AdS_5$ Gaiotto-Maldacena backgrounds and some highly supersymmetric $AdS_4$ and $AdS_3$ backgrounds. For the $AdS_7$ and $AdS_5$ cases our results are consistent with the findings of \cite{Nunez:2018ags,Nunez:2018qcj}. For one of the $AdS_3$ backgrounds we find the eigenvalue $a=1$ which does not rule out factorization of scattering. However, we show in appendix \ref{app:F4} that looking at another 2-to-2 amplitude rules out factorization of scattering also for this background.

The plan of the paper is as follows. In section \ref{sec:lc-gauge} we describe the light-cone gauge bosonic string in the null cusp background. We then derive, in section \ref{sec:constraint}, the constraint on the warp factor from factorization of tree-level scattering on the GKP/null cusp string worldsheet. In section \ref{sec:non-int} we apply this condition to various known supersymmetric $AdS$ backgrounds and show that all but one fail to satisfy it. We end with some conclusions. Appendix \ref{app:remaining} contains a calculation of the remaining 2-to-2 amplitudes that are required to vanish and, as a consistency check, we show that this does not lead to any additional conditions, while appendix \ref{app:F4} contains the additional calculations needed to rule out factorization of scattering for the $AdS_3$ background with $a=1$.

In the standalone appendix \ref{app:sym-NSNS} we consider (unwarped) symmetric space backgrounds for which integrability has been argued, using the factorization of certain massless amplitudes, to lead to certain conditions on the NSNS flux \cite{Wulff:2017hzy,Wulff:2017vhv}. We show that the same conditions are obtained by requiring amplitudes with only two massless particles in the in- and out-states and two massive particles in the out-state to vanish, lending further support to the classification of integrable symmetric space backgrounds in \cite{Wulff:2017vhv}.

\section{Light-cone gauge string in null cusp background}\label{sec:lc-gauge}
The metric is that of a warped product of $AdS_n$ ($n>2$) with an internal space $M_{10-n}$
\begin{equation}
ds^2=e^{2A}ds^2_{AdS_n}+ds^2_{M_{10-n}}\,.
\end{equation}
The metric of $M_{10-n}$ is
\begin{equation}
ds^2_{M_{10-n}}=g_{ij}(x)dx^idx^j
\end{equation}
and the warp factor depends on the position in the internal space $A=A(x)$. We will take the $AdS$ metric to be of the form
\begin{equation}
ds^2_{AdS_n}=e^{-2\varphi}(dz^+dz^-+dz_m^2)+d\varphi^2
\end{equation}
where $m=1,\ldots,n-3$ labels the transverse $AdS$ boundary coordinates and $\varphi$ the radial coordinate. We will also assume there is no NSNS B-field on $AdS$, which is only a restriction for $AdS_3$ backgrounds.

In the bosonic string sigma model we gauge fix the worldsheet metric to the form $\sqrt{-h}h^{ij}=\mathrm{diag}(-e^{2\varphi},e^{-2\varphi})$ and the null cusp solution then takes the form \cite{Giombi:2009gd}
\begin{equation}
e^{2\varphi}=\frac{\tau}{\sigma}\,,\qquad
z^+=\tau\,,\qquad
z^-=-\frac{1}{2\sigma}\,,
\label{eq:ncsol}
\end{equation}
with all other fields vanishing. Fixing light-cone gauge, $z^+=\tau$, and expanding the action we find, upon redefining $z_m\rightarrow\sqrt{\tau/\sigma}z_m$, taking $\tau\rightarrow i\tau$ and changing worldsheet coordinates to $\tau=e^t$ and $\sigma=e^s$,
\begin{align}
g^{-1}L_{\mathrm{nc}}=&
-\tfrac12
e^{2A(x)}
\left[
ie^s\dot z^-
+(z_m/2+\dot z_m)^2
+e^{-4\varphi}(z_m/2-z'_m)^2
+e^{2\varphi}(\tfrac12+\dot\varphi)^2
+e^{-2\varphi}(\tfrac12-\varphi')^2
\right]
\nonumber\\
&{}
-\tfrac12e^{2\varphi}g_{ij}(x)\dot x^i\dot x^j
-\tfrac12e^{-2\varphi}g_{ij}(x)x'^ix'^j
-B_{ij}(x)\dot x^ix'^j\,.
\label{eq:Lnc}
\end{align}
Here all fields denote the \emph{fluctuations} around the null cusp solution with a dot(prime) denoting $\partial_t(\partial_s)$ and the coupling $g$ the string tension. $B_{ij}$ is the B-field (which we allow only on the internal space). Note that the null cusp solution (\ref{eq:ncsol}) is a consistent solution only if the warp factor has no linear term in $x$, so we must assume this to be the case. Note also the $\dot z^-$-term which is a total derivative in the unwarped case. Here it will contribute to interactions involving the non-constant piece of the warp factor. These interaction terms are found by substituting for $ie^s\dot z^-$ the solution of the Virasoro constraints
\begin{align}
ie^s\dot z^-
=&
-(z_m/2+\dot z_m)^2
+e^{-4\varphi}(z_m/2-z'_m)^2
-e^{2\varphi}(\tfrac12+\dot\varphi)^2
+e^{-2\varphi}(\tfrac12-\varphi')^2
\nonumber\\
&{}
-e^{-2(A-\varphi)}g_{ij}(x)\dot x^i\dot x^j
+e^{-2(A+\varphi)}g_{ij}(x)x'^ix'^j\,.
\label{eq:zdot}
\end{align}
In the next section we will carry out the expansion of the action to the required order to compute the amplitude for tree-level $zz\rightarrow xx$ scattering.

\section{Constraint on warp-factor from factorization of scattering}\label{sec:constraint}
Expanding around a (regular) critical point of the warp factor, so that a GKP/null cusp solution exists, we have (rescaling the coordinates suitably)
\begin{equation}
e^{2A}=1+2a_{ij}x^ix^j+\mathcal O(x^3)\,,\qquad g_{ij}=\delta_{ij}+\mathcal O(x)\,.
\label{eq:eA}
\end{equation}
We can diagonalize the symmetric matrix $a_{ij}$ by a rotation and replace it by $\mathrm{diag}(a_i)$.

Rescaling the fields in (\ref{eq:Lnc}) to have canonical kinetic terms and expanding in the coupling $g^{-1/2}$ we find the quadratic action (dropping total derivatives and a constant term)
\begin{equation}
L_2=
-\tfrac12(\partial z_m)^2
-\tfrac14z_m^2
-\tfrac12(\partial\varphi)^2
-\tfrac12\varphi^2
-\tfrac12(\partial x_i)^2
-\tfrac12a_ix_i^2
\end{equation}
and we read off the spectrum\footnote{Often a normalization is chosen where the masses are $\sqrt2$ and $2$ instead of $1/\sqrt2$ and $1$ as here.}
\begin{center}
\begin{tabular}{c|c}
field & mass\\
\hline $z_m$ & $\tfrac{1}{\sqrt2}$\\
$\varphi$  & $1$\\
$x_i$ & $\sqrt{a_i}$
\end{tabular}
\end{center}

We will now write the cubic and quartic interaction terms, but since it will be enough for our purposes to retain only one of the $z_m$ and one of the $x_i$, say $z_1$ and $x_1$, we will drop the indices $m,i$ in the following. The cubic interaction terms are
\begin{equation}
g^{1/2}L_3=
2\varphi(z/2-z')^2
+ax^2(\varphi+2\varphi')
-\varphi(\dot x^2-x'^2)
-\varphi(\dot\varphi^2-\varphi'^2)
+(x^3-\mbox{terms})\,,
\label{eq:L3}
\end{equation}
while the quartic terms are
\begin{align}
gL_4=&
-2ax^2(z/2-z')^2
-4\varphi^2(z/2-z')^2
-2ax^2(\varphi+\varphi')^2
+ax^2\varphi^2
\nonumber\\
&{}
-\varphi^2(\partial x)^2
-\varphi^2(\partial\varphi)^2
-\tfrac16\varphi^4
+(\varphi x^3-\mbox{terms})
+(x^4-\mbox{terms})
\label{eq:L4}
\end{align}
Note that to derive these we have used (\ref{eq:zdot}).

To be consistent with factorized scattering all amplitudes involving particles of one mass going into particles of different mass must vanish. In particular, recalling that we have particles ($z,\,\varphi,\,x$) with $m^2=(\frac12,\,1,\,a)$ respectively, the amplitudes for $zz\rightarrow\varphi\varphi$, $zz\rightarrow xx$ and $xx\rightarrow\varphi\varphi$ must all vanish unless either $a=\frac12$ or $a=1$ in which case respectively the second or third amplitude is not constrained. To derive the constraints on $a$ it turns out to be sufficient to consider the amplitude for $zz\rightarrow xx$. This is true for all cases except $AdS_3$ since in that case the $z$ excitations are absent. In appendix \ref{app:remaining} we show, for completeness and as a consistency check, that the vanishing of the $zz\rightarrow\varphi\varphi$ and $xx\rightarrow\varphi\varphi$ amplitudes leads to the same constraints. The only exception is $AdS_3$ where we cannot rule out the additional possibility $a=1$ since in that case all the excitations have the same mass and there are no constraints.

The amplitude for $zz\rightarrow xx$ has two contributions at tree-level. A contact term coming from the $zzxx$-vertex in (\ref{eq:L4}) and an s-channel contribution from the $zz\varphi$ and $xx\varphi$ vertices in (\ref{eq:L3}). Since there are no $zzx$ or $zxx$ terms in (\ref{eq:L3}) there are no t or u-channel contributions. We will use light-cone momenta, which in the Euclidean case are $p_\pm=-ie\pm p$. The contact diagram gives
\begin{equation}
\mathcal A_c(zz\rightarrow xx)=-2ag^{-1}A\,,\qquad A=(1+ip_{1+}-ip_{1-})(1+ip_{2+}-ip_{2-})\,,
\end{equation}
while the s-channel contribution is
\begin{equation}
\mathcal A_s(zz\rightarrow xx)
=
ag^{-1}A(1-2p_{1+}p_{2+})
\frac{1-2i(p_{1+}+p_{2+})-2p_{1+}p_{2+}}{p_{1+}^2+p_{2+}^2}\,,
\end{equation}
where we have used the fact that $p_{3+}p_{4+}=2ap_{1+}p_{2+}$. Clearly the total amplitude can only vanish if $a=0$. Therefore factorization of scattering requires either $a=0$, so that the amplitude vanishes, or $a=\frac12$ so that the masses of incoming and outgoing particles are the same and the amplitude is not required to vanish to be compatible with factorization of scattering.

We can repeat the same computation for each $x_i$ therefore we find that integrability requires the eigenvalues of $a_{ij}$ in (\ref{eq:eA}) to satisfy $a_i=\{0,\frac12\}$. We will now use this constraint to rule out integrability for various known supersymmetric $AdS$ backgrounds.

\section{Non-integrability of supersymmetric $AdS$ backgrounds}\label{sec:non-int}

\subsection{$AdS_7$ backgrounds}
The equations for the most general supersymmetric type II $AdS_7$ solutions were found in \cite{Apruzzi:2013yva} (see also \cite{Gaiotto:2014lca} for the AdS/CFT interpretation of these backgrounds). There are no solutions in type IIB. For the type IIA solutions the metric takes the form (\ref{eq:warped-metric}) with the internal space metric being (eq. (4.16) in \cite{Apruzzi:2013yva}, we use upper case $X$ rather than their $x$)
\begin{equation}
ds^2_{M_3}=dr^2+\frac{1}{16}e^{2A}(1-X^2)ds^2_{S^2}\,.
\end{equation}
Here the warp factor $A$, the dilaton $\phi$ and the function $X$ ($|X|\leq1$) are functions of $r$ satisfying the system of ODE's (eq. (4.17) in \cite{Apruzzi:2013yva})
\begin{align}
\phi'=&\,{}\frac14\frac{e^{-A}}{\sqrt{1-X^2}}\left(12X+(2X^2-5)F_0e^{A+\phi}\right)\,,\\
X'=&\,{}-\frac12e^{-A}\sqrt{1-X^2}\left(4+XF_0e^{A+\phi}\right)\,,\\
A'=&\,{}\frac14\frac{e^{-A}}{\sqrt{1-X^2}}\left(4X-F_0e^{A+\phi}\right)\,,
\end{align}
where the prime denotes differentiation with respect to $r$ and the constant $F_0$ is the Romans mass parameter or RR $0$-form field strength. We will not need the expressions for the fluxes here.

Expanding around a critical point $r=r_0$, $A'|_{r=r_0}=0$,\footnote{If we take $X_0=\pm1$ we are then sitting at the "North Pole"("South Pole") at the ends of the interval where $r$ is defined \cite{Gaiotto:2014lca}.} so that the null cusp solution exists, one finds
\begin{equation}
\phi=\phi_0+\phi_1r+\mathcal O(r^2)\,,\qquad
X=X_0+X_1r+\mathcal O(r^2)\,,\qquad
A=A_0+A_2r^2+\mathcal O(r^3)\,,
\end{equation}
with
\begin{equation}
X_0=\frac14F_0e^{A_0+\phi_0}\,,\quad
X_1=-2e^{-A_0}(1+X_0^2)\sqrt{1-X_0^2}\,,\quad
\phi_1=-2e^{-A_0}X_0\sqrt{1-X_0^2}\,,\quad
A_2=-e^{-2A_0}\,.
\end{equation}
The relevant part of the metric is
\begin{equation}
ds^2=e^{2A_0}(1+2A_2r^2+\mathcal O(r^3))ds^2_{AdS_7}+dr^2+\ldots\,.
\end{equation}
Comparing to (\ref{eq:eA}) we read off (after rescaling $r$ appropriately) that $a=A_2e^{2A_0}=-1$ (the sign means that the $r$ excitation is tachyonic in this case). Since $a\neq0,\frac12$ we conclude that scattering fails to factorize in this background.

\subsection{$AdS_6$ backgrounds}
The most general solution in type IIA is the Brandhuber-Oz solution \cite{Brandhuber:1999np}, as was shown in \cite{Passias:2012vp}. The metric is of the form (\ref{eq:warped-metric}) with
\begin{equation}
e^{2A}=\frac94\left(\frac32F_0\sin\alpha\right)^{-1/3}\,,\qquad
ds^2_{M_4}=\frac49e^{2A}(d\alpha^2+\cos^2\alpha\,ds^2_{S^3})\,.
\end{equation}
To avoid a linear term in the warp factor we should expand around the point $\alpha=\pi/2$. Doing this the relevant part of the metric becomes
\begin{equation}
e^{-2A_0}ds^2=\left(1+\frac16\alpha^2+\mathcal O(\alpha^4)\right)ds^2_{AdS_6}+\frac49d\alpha^2+\ldots\,,\qquad e^{2A_0}=\frac94\left(\frac32F_0\right)^{-1/3}\,.
\end{equation}
Comparing to (\ref{eq:eA}) we read off (after rescaling $\alpha$ appropriately) that $a=\frac{3}{16}$ which rules out factorization of scattering for this background. A small comment is in order here. Around the point we are expanding the $S^4$ part of the metric looks like
\begin{equation}
d\alpha^2+\alpha^2ds^2_{S^3}\,.
\end{equation}
We can consistently truncate the $S^3$ modes but if we want to keep them we need to change coordinates to have a regular perturbative expansion of the sigma model around $\alpha=0$. We then go from spherical to Cartesian coordinates with metric $dx_i^2$ where now $\alpha=\sqrt{x_i^2}$. Focusing now on one of the $x_i$ excitations we find the same mass that we found for $\alpha$ above, so the conclusion is the same, as it should be. Similar comments apply to some of the backgrounds considered below.

The equations for the most general supersymmetric type IIB $AdS_6$ solutions were determined in \cite{Apruzzi:2014qva} (see also for example \cite{Kim:2015hya,DHoker:2016ujz}). The metric takes the form (\ref{eq:warped-metric}) with (eq. (4.16) in \cite{Apruzzi:2014qva})
\begin{equation}
ds^2_{M_4}=\frac{\cos\alpha}{\sin^2\alpha}\frac{dq^2}{q}+\frac19q(1-x^2)\frac{\sin^2\alpha}{\cos\alpha}\left[\frac{1}{x^2}\left(\frac{dp}{p}+3\cot^2\alpha\frac{dq}{q}\right)^2+ds^2_{S^2}\right]\,,
\end{equation}
where $q=e^{2A}\cos\alpha$ and $p=e^{4A-\phi}\sin\alpha\sqrt{1-x^2}$. Here the warp factor and dilaton are functions of $x$ and $\alpha$ subject to the PDEs (eqs. (5.1a), (5.1b) in \cite{Apruzzi:2014qva})
\begin{align}
3\sin(2\alpha)(\dot A\phi'-A'\dot\phi)=&\,{}6A'+\sin^2\alpha(-2x-2(x^2+5)A'+(1+2x^2)\phi')\,,\label{eq:AdS6-1}\\
\cos\alpha(2+3x\phi')+\sin\alpha\dot\phi=&\,{}2x\left(\frac{3}{\sin\alpha}+(x^2-4)\sin\alpha\right)(\dot A\phi'-A'\dot\phi)
\label{eq:AdS6-2}
\\&\,{}
-2x\cos\alpha\left(\frac{3}{\sin^2\alpha}-(x^2+5)\right)A'+2\left(\frac{3}{\sin\alpha}-(x^2+1)\sin\alpha\right)\dot A\,,
\nonumber
\end{align}
where a dot denotes differentiation with respect to $\alpha$ and a prime differentiation with respect to $x$. Since the case $x=0$ gives the T-dual of the Branhuber-Oz solution \cite{Apruzzi:2014qva} we take $0<x\leq1$.

Expanding around a point $(\alpha_0,x_0)$ with $\dot A(\alpha_0,x_0)=A'(\alpha_0,x_0)=0$ we get, assuming $\sin(2\alpha_0)\neq0$,
\begin{equation}
\phi'|=2\frac{x_0}{1+2x_0^2}\,,\qquad
\dot\phi|=-2\cot\alpha_0\frac{1+5x_0^2}{1+2x_0^2}\,.
\end{equation}
Using this we find that the metric is diagonal in $(\alpha,y)$ with $y=x-2x_0(1-x_0^2)\cot\alpha_0\alpha$. Writing $e^{2A}=e^{2A_0}(1+2a_{ij}x^ix^j+\mathcal O(x^3))$, we diagonalize $a_{ij}$ which implies that $\partial_\alpha\partial_yA|=0$ or
\begin{equation}
\dot A'|=-2x_0(1-x_0^2)\cot\alpha_0A''|
\label{eq:Adotprime}
\end{equation}
and
\begin{equation}
e^{2A}=e^{2A_0}(1+2a_1\alpha^2+2a_2y^2+\mathcal O(x^3))\,,\qquad\ddot A|=2a_1+8a_2x_0^2(1-x_0^2)^2\cot^2\alpha_0\,,\qquad A''|=2a_2\,.
\label{eq:AdS6-e2A}
\end{equation}
Taking the $\alpha$ and $x$-derivative of (\ref{eq:AdS6-1}) lets us solve for $\dot\phi'$ and $\phi''$. Plugging the result into the $x$-derivative of (\ref{eq:AdS6-2}) gives, after dividing by $x_0\cos\alpha_0$,
\begin{equation}
1
+(1-x_0^2)\left(4x_0^2(1-x_0^2)+\frac{1+8x_0^2}{\sin^2\alpha_0}\right)A''|
+4x_0(1-x_0^2)\cot\alpha_0\dot A'|
+\ddot A|
=
0
\end{equation}
or, using (\ref{eq:Adotprime}) and (\ref{eq:AdS6-e2A}),
\begin{equation}
1
+2(1-x_0^2)\left(8x_0^2(1-x_0^2)+\frac{(1+2x_0^2)^2}{\sin^2\alpha_0}\right)a_2
+2a_1
=
0
\end{equation}
and since the coefficient of $a_2$ is positive this implies that either $a_2<0$ or $a_1<0$. The relevant part of the metric is
\begin{equation}
e^{-2A_0}ds^2=\left(1+2a_1\alpha^2+2a_2y^2+\ldots\right)ds^2_{AdS_6}
+d\alpha^2
+\frac{\sin^2\alpha_0 dy^2}{(1-x_0^2)(1+2x_0^2)^2}
+\ldots
\end{equation}
and from this it is easy to see that if $a_2<0$ the $y$-excitation is tachyonic and if $a_2\geq0$ the $\alpha$-excitation is tachyonic. Neither is compatible with factorization of scattering.

\subsection{$AdS_5$ Gaiotto-Maldacena backgrounds}
Eleven-dimensional supergravity backgrounds dual to four-dimensional $\mathcal N=2$ SCFTs were constructed in \cite{Gaiotto:2009gz}. Here we will consider the solutions with a $U(1)$-isometry which can be used to reduce to type IIA supergravity \cite{ReidEdwards:2010qs}. In that case the solutions are in correspondence with solutions to the axially symmetric three-dimensional Laplace equation
\begin{equation}
\frac1\rho\partial_\rho(\rho\partial_\rho V)+\partial^2_\eta V=0\,,
\end{equation}
subject to certain boundary conditions. The only information about the solutions relevant for us is that the metric takes the form
\begin{align}
ds^2=&\,2(\rho^2+2f)^{1/2}\Big(2ds^2_{AdS_5}+f^{-1}(d\eta^2+d\rho^2)
+f\left(\rho^2+2f+f^2(\partial_\eta\partial_\rho V/\partial_\rho V)^2\right)^{-1}ds^2_{S^2}
\nonumber\\
&\qquad\qquad\qquad{}
+2(\rho^2+2̇f)^{-1}d\chi^2
\Big)\qquad\qquad\mbox{with}\qquad f=\frac{\rho\partial_\rho V}{\partial_\eta^2V}\,.
\label{eq:dsGM}
\end{align}
The basic example of such a solution is the Maldacena-N\'u\~nez solution \cite{Maldacena:2000mw} for which
\begin{equation}
V_{MN}=\frac12\sqrt{\rho^2+(N+\eta)^2}-\frac12(N+\eta)\sinh^{-1}\left(\frac{N+\eta}{\rho}\right)-(\eta\leftrightarrow-\eta)\,.
\end{equation}
In fact it was shown in \cite{ReidEdwards:2010qs} that the most general solution satisfying certain boundary conditions can be expressed as a linear combination of Maldacena-N\'u\~nez solutions.\footnote{It was shown in \cite{Aharony:2012tz} that the boundary conditions can be relaxed to describe solutions with more than one stack of NS5-branes. We will not consider this more general case here.}

We will start by analyzing the Maldacena-N\'u\~nez solutions for which the function appearing in the metric takes the form
\begin{equation}
f_{MN}=(\rho^2+(N+\eta)^2)^{1/2}(\rho^2+(N-\eta)^2)^{1/2}\,.
\end{equation}
Expanding to linear order in $\eta$ and quadratic order in $x=\rho/(\sqrt2N)$ we get
\begin{equation}
\frac{\sqrt2}{8N}ds^2=\left(1+\frac{3}{2}x^2+\ldots\right)ds^2_{AdS_5}+dx^2+\ldots\,.
\end{equation}
Comparing to (\ref{eq:eA}) we read off $a=\frac34$ which is not compatible with factorized scattering.

We now turn to the general solutions. There are two classes referred to as periodic and non-periodic. The general non-periodic solution can be written as sum of Maldacena-N\'u\~nez solutions with $N=m_i$ \cite{ReidEdwards:2010qs}
\begin{equation}
V_{\mathrm{Non-periodic}}=\sum_{i=1}^nc_iV_{MN}(m_i)\,,
\end{equation}
with positive coefficients, $c_i\geq0$. In this case we have from (\ref{eq:dsGM})
\begin{equation}
f=
\frac{\sum_{i=1}^nc_i\left[(\rho^2+(m_i+\eta)^2)^{1/2}-(\rho^2+(m_i-\eta)^2)^{1/2}\right]}
{
\sum_{i=1}^nc_i\left[(\rho^2+(m_i-\eta)^2)^{-1/2}-(\rho^2+(m_i+\eta)^2)^{-1/2}\right]
}\,.
\end{equation}
Expanding around $\eta=0$ we get
\begin{equation}
f=
\frac{\sum_{i=1}^nc_im_i(\rho^2+m_i^2)^{-1/2}}{\sum_{i=1}^nc_im_i(\rho^2+m_i^2)^{-3/2}}
+\mathcal O(\eta^2)
=
\frac{s_0}{s_2}\left(1+\frac{3s_4}{2s_2}\rho^2-\frac{s_2}{2s_0}\rho^2\right)
+\mathcal O(\eta^2,\rho^4)\,,
\end{equation}
where we have defined the sums $s_k=\sum_{i=1}^n\frac{c_i}{m_i^k}$. Writing $\rho=\sqrt{\frac{2s_0}{s_2}}\,x$ this gives for the metric
\begin{equation}
\frac14\left(\frac{s_2}{2s_0}\right)^{1/2}ds^2=\left(1+\frac{3s_0s_4}{2s_2^2}x^2+\ldots\right)ds^2_{AdS_5}+dx^2+\ldots
%
\end{equation}
We now make use of a special case of H\"older's inequality which takes the form
\begin{equation}
\sum_{i=1}^n|x_iy_i|\leq\left(\sum_{i=1}^n|x_i|^2\right)^{1/2}\left(\sum_{i=1}^n|y_i|^2\right)^{1/2}\,.
\end{equation}
Taking $x_i=\sqrt{c_i}$ and $y_i=\sqrt{c_i}\,m_i^{-2}$ and squaring gives $s_2^2\leq s_0s_4$ so that, comparing to (\ref{eq:eA}) we find that $a\geq\frac34$ ruling out integrability.

It remains to analyze the general periodic solution for which we now have an infinite sum \cite{ReidEdwards:2010qs}
\begin{equation}
V_{\mathrm{Periodic}}=\sum_{\alpha=-\infty}^\infty\sum_{i=1}^nc_iV_{MN}(\rho,\alpha\Lambda-\eta;m_i)\,,
\end{equation}
where $\Lambda=2\sum_{i=1}^nm_i$. The function appearing in the metric (\ref{eq:dsGM}) now becomes
\begin{equation}
f=
\frac{\sum_{\alpha=-\infty}^\infty\sum_{i=1}^nc_i\left[(\rho^2+(m_i+\alpha\Lambda-\eta)^2)^{1/2}-(\rho^2+(m_i-\alpha\Lambda+\eta)^2)^{1/2}\right]}
{\sum_{\alpha=-\infty}^\infty\sum_{i=1}^nc_i\left[(\rho^2+(m_i-\alpha\Lambda+\eta)^2)^{-1/2}-(\rho^2+(m_i+\alpha\Lambda-\eta)^2)^{-1/2}\right]}\,.
\end{equation}
Expanding around $\eta=0$ we get
\begin{align}
f=&\,
\frac{\sum_{\alpha=-\infty}^\infty\sum_{i=1}^nc_i[(m_i+\alpha\Lambda)(\rho^2+(m_i+\alpha\Lambda)^2)^{-1/2}+(m_i-\alpha\Lambda)(\rho^2+(m_i-\alpha\Lambda)^2)^{-1/2}]}
{\sum_{\alpha=-\infty}^\infty\sum_{i=1}^nc_i[(m_i+\alpha\Lambda)(\rho^2+(m_i+\alpha\Lambda)^2)^{-3/2}+(m_i-\alpha\Lambda)(\rho^2+(m_i-\alpha\Lambda)^2)^{-3/2}]}
+\mathcal O(\eta^2)
\nonumber\\
=&\,
\frac{S_0}{S_2}\left(1+\frac{3S_4}{2S_2}\rho^2-\frac{S_2}{2S_0}\rho^2\right)
+\mathcal O(\eta^2,\rho^4)\,,
\end{align}
which is of the same form as before except that now 
\begin{align}
S_k=&\,
\frac12\sum_{\alpha=-\infty}^\infty\sum_{i=1}^nc_i[\mathrm{sign}(m_i+\alpha\Lambda)(m_i+\alpha\Lambda)^{-k}+\mathrm{sign}(m_i-\alpha\Lambda)(m_i-\alpha\Lambda)^{-k}]
\nonumber\\
=&\,
s_k
+\sum_{\alpha=1}^\infty\sum_{i=1}^nc_i[(m_i+\alpha\Lambda)^{-k}-(m_i-\alpha\Lambda)^{-k}]\,.
\end{align}
Therefore we now find $a=\frac{3S_0S_4}{4S_2^2}$. We have $S_0=s_0$ and $S_2=s_2-r_2$ and $S_2=s_4-r_4$ with $r_2,r_4>0$ (of course also $S_k>0$). We then have
\begin{equation}
\frac{S_0S_4}{S_2^2}
=
\frac{s_0(s_4-r_4)}{(s_2-r_2)^2}
\geq
\frac{s_0(s_4-r_4)}{s_2(s_2-r_2)}
\geq
\frac{1-r_4/s_4}{1-r_2/s_2}
\geq
\frac{1-r_4/s_4}{1-r_2/s_4}\,,
\end{equation}
where in the third step we used the fact that $s_0s_4/s_2^2\geq1$ derived above. Finally we note that\footnote{This follows from
$$
(m_i-\alpha\Lambda)^{-4}-(m_i+\alpha\Lambda)^{-4}
=
[(m_i-\alpha\Lambda)^{-2}-(m_i+\alpha\Lambda)^{-2}][(m_i+\alpha\Lambda)^{-2}+(m_i-\alpha\Lambda)^{-2}]
\leq
(m_i-\alpha\Lambda)^{-2}-(m_i+\alpha\Lambda)^{-2}\,,
$$
where we used the fact that $\alpha$ and $m_i$ are integers, recall that $\Lambda=2\sum_{i=1}^nm_i$. 
} $r_2\geq r_4$ which means that $a=\frac{3S_0S_4}{4S_2^2}\geq\frac34$ which is not compatible with factorization of scattering.

\subsection{$AdS_4$ background preserving $\mathcal N=3$ supersymmetry}
We will now consider the $AdS_4$ solution of massive type IIA preserving $\mathcal N=3$ supersymmetry presented in \cite{Pang:2015vna,DeLuca:2018buk}, obtained by uplifting the $D=4$ solution constructed in \cite{Gallerati:2014xra}. Converting to string frame the relevant part of the metric is (eq. (5.1) of \cite{DeLuca:2018buk})
\begin{equation}
ds^2=(3+\cos2\alpha)^{1/2}\left[ds^2_{AdS_4}+2d\alpha^2+\ldots\right]\,.
\end{equation}
Expanding around $\alpha=0$ and writing $\alpha=x/\sqrt2$ we have
\begin{equation}
\frac12ds^2=\left(1-\frac18x^2+\ldots\right)ds^2_{AdS_4}+dx^2+\ldots
\end{equation}
and we read off $a=-\frac{1}{16}$ which is again not compatible with factorization of scattering.

\subsection{$AdS_3$ backgrounds with exceptional superisometry groups $F(4)$ and $G(3)$}
In \cite{Dibitetto:2018ftj} type IIA $AdS_3$ solutions preserving $\mathcal N=7,8$ supersymmetry with exceptional superisometry groups $F(4)$ and $G(3)$ were constructed. The metric of the $F(4)$ solution is (eq. (4.7) in \cite{Dibitetto:2018ftj})
\begin{equation}
R^{-2}ds^2
=
\frac{1+z^3}{\sqrt z}ds^2_{AdS_3}
+\frac94\frac{\sqrt z}{1+z^3}dz^2
+\frac94\frac{1}{\sqrt z}ds^2_{S^6}\,.
\label{eq:gF4}
\end{equation}
Expanding around $z=z_0=5^{-1/3}$ the warp factor has no linear term and a GKP solution exists. Writing $z=z_0(1+4x/\sqrt5)$ we get
\begin{equation}
\frac{5^{5/6}}{6}R^{-2}ds^2
=
\left(1+2x^2+\mathcal O(x^3)\right)ds^2_{AdS_3}
+dx^2
+\ldots
\,,
\end{equation}
from which we read off the mass square of $x$ as $a=1$ which is compatible with factorization of scattering since the constraints are weaker in $AdS_3$. Therefore our analysis for this background is not conclusive. However one can rule out integrability for this background by showing that the amplitude for two massless excitations from the $S^6$ going into two $x$ excitations of mass 1 is non-zero, the calculations are presented in appendix \ref{app:F4}.

The metric of the $G(3)$ solution is (eq. (5.14a) in \cite{Dibitetto:2018ftj})
\begin{equation}
R^{-2}ds^2=\frac{(1+9y^2)(1+y^2)^{1/3}}{4y^{5/3}}ds^2_{AdS_3}+\frac{4}{9(1+9y^2)(1+y^2)^{5/3}y^{5/3}}dy^2+\frac{y^{1/3}}{(1+y^2)^{2/3}}ds^2_{S^6}\,.
\end{equation}
Expanding around $y=y_0=3^{-1/2}$ the warp factor has no linear term and a GKP solution exists. Writing $y=y_0+4x$ we find
\begin{equation}
3^{-1/2}4^{-1/3}R^{-2}ds^2=\left(1+24x^2+\mathcal O(x^3)\right)ds^2_{AdS_3}+dx^2+\ldots\,,
\end{equation}
from which we read off $a=12$, again incompatible with factorized scattering.

\section{Conclusions}
We have shown that the requirement of factorization of scattering on the worldsheet of a GKP spinning string forces an $AdS$ background to look, locally, to quadratic order in the relevant excitation, like an unwarped background (with the possible exception of $AdS_3$ where $a=1$ is allowed). This constraint is surprisingly strong and already rules out integrability for many highly supersymmetric $AdS$ backgrounds known in the literature. The non-integrability of the $AdS_7$ and $AdS_5$ Gaiotto-Maldacena backgrounds was previously argued in \cite{Nunez:2018ags} and \cite{Nunez:2018qcj} using the more sophisticated tools of analytic integrability. Our findings are in agreement with the findings of those papers.\footnote{In \cite{Filippas:2019puw} it was shown that a certain supersymmetric $AdS_7$ background leads to an integrable bosonic string. However that background does not solve the standard type IIA supergravity equations, but rather the equations with D8-brane sources. In fact the solution is such that the source has support everywhere. It is not clear how to make sense of the superstring in such a background since for example kappa symmetry of the Green-Schwarz string implies the supergravity equations without sources, e.g. \cite{Wulff:2016tju}.} In addition we ruled out integrability for supersymmetric $AdS_6$ backgrounds under the assumption that they allow for a regular critical point for the warp factor, which is necessary to have a GKP string solution. We also saw that some special, highly supersymmetric, $AdS_4$ and $AdS_3$ backgrounds also fail to be integrable.

Since we only need the expansion of the warp factor to second order in fluctuations around a regular critical point to apply the condition derived here it should be possible to rule out integrability of backgrounds that are only known locally in perturbation theory. In this way it should be possible to look for integrability of backgrounds that are not known explicitly, or like some of the examples here, reduced only to solving a few ODEs. The question is if the backgrounds are constrained enough by the supergravity equations that the condition on the warp factor becomes strong. We hope to return to this question in the future.

Note that we have assumed here that the $AdS$ isometries are not broken. There are also interesting backgrounds where they are broken, a recent example being the background constructed in \cite{Bobev:2019ylk}, which is related to the omega-deformation of the gauge theory. In that background the $AdS$ isometries are broken by the B-field (setting their $X=1$)
\begin{equation}
B\sim e^{-\phi}\sin\theta\,e^{-\varphi}(dz^+\wedge dz_1+dz^-\wedge dz_2)\,,
\end{equation}
where $\phi$ and $\theta$ are angles on the $S^5$, while the metric is undeformed. The null-cusp/GKP string solution is still consistent around $\phi=\theta=0$. However, already at quadratic order the action will look unconventional involving terms like $\theta\partial_\sigma z_1$. Presumably a field redefinition can be used to get rid of these terms but the analysis will require some extra work which is beyond the scope of the present paper.

\vspace{1cm}

\section*{Acknowledgments}
I am thankful to B. Hoare, N. Levine and A. Tseytlin for interesting discussions and helpful comments on the draft.

\vspace{1cm}

\appendix
\section{Amplitude for $zz\rightarrow\varphi\varphi$ and $xx\rightarrow\varphi\varphi$}\label{app:remaining}
Here we show that the vanishing of the remaining tree-level two-particle amplitudes not considered in section \ref{sec:constraint}, $zz\rightarrow\varphi\varphi$ and $xx\rightarrow\varphi\varphi$, do not lead to additional constraints. Except in the case of $AdS_3$ where there is no $z$ excitation, this is just a consistency check since it is clear from the arguments in the introduction that there cannot be further constraints on the warp factor.

We start with the amplitude for $zz\rightarrow\varphi\varphi$. The contact contribution is
\begin{equation}
\mathcal A_c(zz\rightarrow\varphi\varphi)
=
-4A\,,\qquad
A=g^{-1}(1+ip_{1+}-ip_{1-})(1+ip_{2+}-ip_{2-})\,.
\end{equation}
The s-channel contribution is
\begin{equation}
\mathcal A_s(zz\rightarrow\varphi\varphi)
=
A
\frac{
p_{3+}p_{4+}
+p_{3-}p_{4-}
-(p_{3+}+p_{4+})^2
-(p_{3-}+p_{4-})^2
}
{(p_{1+}+p_{2+})(p_{1-}+p_{2-})-1}
=
-2A
\frac{
p_{1+}^2p_{2+}^2
+1/4
}
{p_{1+}p_{2+}}\,,
\end{equation}
where we used the fact that for the relevant kinematics $p_{3+}p_{4+}=2p_{1+}p_{2+}$. Finally the t- and u-channel diagrams give
\begin{equation}
\mathcal A_{t+u}(zz\rightarrow\varphi\varphi)
=
-A
\frac{1+(p_{1+}-p_{3+}-p_{1-}+p_{3-})^2}{(p_{1+}-p_{3+})(p_{1-}-p_{3-})-1/2}
+(p_3\leftrightarrow p_4)
=
A
\frac{
1
+8p_{1+}p_{2+}
+4p_{1+}^2p_{2+}^2
}{2p_{1+}p_{2+}}\,,
%
\end{equation}
where we used the useful identity $(p_{1\pm}-p_{3\pm})(p_{1\pm}-p_{4\pm})=p_{1\pm}p_{2\pm}$ and similar identities. Summing these up we find that the total amplitude vanishes as expected.

Next we turn to the amplitude for $xx\rightarrow\varphi\varphi$ which is the most involved. The contact diagram contribution is
\begin{align}
\mathcal A_c(xx\rightarrow\varphi\varphi)
=&\,
2g^{-1}
\Big[
-(p_{1+}p_{2-}+p_{1-}p_{2+})
-2a
-2ai(p_{3+}+p_{4+})
+2ai(p_{3-}+p_{4-})
\nonumber\\
&{}
+a(p_{3+}-p_{3-})(p_{4+}-p_{4-})
\Big]\,.
%
\end{align}
The s-channel contribution is
\begin{align}
\mathcal A_s(xx\rightarrow\varphi\varphi)
=&\,
g^{-1}
\Big[
2ia([p_{1+}+p_{2+}]-[p_{1-}+p_{2-}])(p_{3+}p_{4+}+p_{3-}p_{4-})
\\
&{}
-2ia([p_{1+}+p_{2+}]^2[p_{3+}+p_{4+}]+[p_{1+}+p_{2+}][p_{1-}+p_{2-}][p_{3-}+p_{4-}])
\nonumber\\
&{}
+2ia([p_{1+}+p_{2+}][p_{1-}+p_{2-}][p_{3+}+p_{4+}]+[p_{1-}+p_{2-}]^2[p_{3-}+p_{4-}])
\nonumber\\
&{}
+(p_{1+}p_{2+}+p_{1-}p_{2-}-2a)([p_{1+}+p_{2+}][p_{3+}+p_{4+}]+[p_{1-}+p_{2-}][p_{3-}+p_{4-}])
\nonumber\\
&{}
-(p_{1+}p_{2+}+p_{1-}p_{2-}-2a)(p_{3+}p_{4+}+p_{3-}p_{4-})
\Big]
/\big[(p_{1+}+p_{2+})(p_{1-}+p_{2-})-1\big]\,.\nonumber
%
\end{align}
Finally the t- and u-channel diagrams give
\begin{align}
\mathcal A_{t+u}(xx\rightarrow\varphi\varphi)
=&\,
g^{-1}
\Big[
(p_{1+}[p_{1+}-p_{3+}]+p_{1-}[p_{1-}-p_{3-}])(p_{2+}[p_{1+}-p_{3+}]+p_{2-}[p_{1-}-p_{3-}])
\nonumber\\
&{}
+2a(ip_{3+}-ip_{3-}+1)[p_{2+}[p_{1+}-p_{3+}]+p_{2-}[p_{1-}-p_{3-}])
\nonumber\\
&{}
-2a(ip_{4+}-ip_{4-}+1)[p_{1+}[p_{1+}-p_{3+}]+p_{1-}[p_{1-}-p_{3-}])
\nonumber\\
&{}
-4a^2(ip_{3+}-ip_{3-}+1)(ip_{4+}-ip_{4-}+1)
\Big]
/\big[(p_{1+}-p_{3+})(p_{1-}-p_{3-})-a\big]
\nonumber\\
&{}
+(p_3\leftrightarrow p_4)
%
%
%
\end{align}

To see that $a=0,\frac12$ is necessary for the total amplitude to vanish it is enough to consider the case $p_{3+}=p_{4+}=(p_{1+}+p_{2+})/2$, in which case we also have $(p_{1+}+p_{2+})^2=4p_{1+}p_{2+}/a$. Looking at the imaginary part of the amplitude only we find
\begin{align}
\mathrm{Im}\,\mathcal A_c=&\,{}-4ag^{-1}(p_{1+}+p_{2+})\left(1-\frac{a}{p_{1+}p_{2+}}\right)\,,\\
\mathrm{Im}\,\mathcal A_s=&\,{}-2g^{-1}(p_{1+}+p_{2+})\left(1-\frac{a}{p_{1+}p_{2+}}\right)p_{1+}p_{2+}\left(1+\frac{a^2}{p_{1+}^2p_{2+}^2}\right)\,,\\
\mathrm{Im}\,\mathcal A_{t+u}=&\,{}4g^{-1}(p_{1+}+p_{2+})\left(1-\frac{a}{p_{1+}p_{2+}}\right)p_{1+}p_{2+}\left([1-a]\left(1+\frac{a^2}{p_{1+}^2p_{2+}^2}\right)+\frac{2a^2}{p_{1+}p_{2+}}\right)\,.
\end{align}
The sum is
\begin{equation}
2(2a-1)g^{-1}(p_{1+}+p_{2+})\left(1-\frac{a}{p_{1+}p_{2+}}\right)\left[2a-p_{1+}p_{2+}\left(1+\frac{a^2}{p_{1+}^2p_{2+}^2}\right)\right]
\end{equation}
and this only vanishes (for general momentum $p_{1+}$) for $a=0,\frac12$. This condition is therefore necessary. It remains to verify that the full amplitude vanishes if $a=0,\frac12$ for general momenta, so that this condition is sufficient.

If $a=0$ the s-channel amplitude vanishes and the t- and u-channel amplitude is easily seen to cancel against the contact amplitude.

For $a=\frac12$ we will analyze the real and imaginary parts of the amplitude separately. Using the fact that $(p_{1\pm}-p_{3\pm})(p_{1\pm}-p_{4\pm})=p_{1\pm}p_{2\pm}$ we find for the imaginary parts exactly the same expressions as above but with $a=\frac12$, which we have already shown to add up to zero. For the real parts we find
\begin{align}
\mathrm{Re}\,\mathcal A_c
=&\,
g^{-1}
\left(
\frac{1}{2p_{1+}p_{2+}}
-\frac{3(p_{1+}^2+p_{2+}^2)}{2p_{1+}p_{2+}}
+2p_{1+}p_{2+}
-1
\right)\,,\\
\mathrm{Re}\,\mathcal A_s=&\,2g^{-1}\left(p_{1+}p_{2+}-\frac12\right)^2\left(1+\frac{1}{4p_{1+}^2p_{2+}^2}\right)\,,\\
\mathrm{Re}\,\mathcal A_{t+u}
=&\,
g^{-1}
\left(
\frac{3(p_{1+}^2+p_{2+}^2)}{2p_{1+}p_{2+}}
-\frac{1}{8p_{1+}^2p_{2+}^2}
-2p_{1+}^2p_{2+}^2
\right)\,,
\end{align}
which also adds up to zero. This completes our consistency check.

\section{Non-factorization for $AdS_3$ background with $F(4)$ isometry}\label{app:F4}
For this background we found $a=1$ and could therefore not rule out integrability. Here we will show that the amplitude for two massless $S^6$ excitations $y$ going into two $x$ excitations of mass 1 is non-zero, ruling out factorization of scattering. Using (\ref{eq:gF4}), expanded around the null cusp solution as $z=5^{-1/3}(1+4x/\sqrt5)$, in (\ref{eq:Lnc}) one finds that the relevant interaction terms are (note that there is no B-field)
\begin{align}
L_{\mathrm{int}}=&{}
g^{-1/2}
\left(
(2\varphi'+\varphi)x^2
-\varphi(\dot x^2-x'^2)
-\varphi(\dot y^2-y'^2)
+\frac{1}{\sqrt5}x(\dot y^2+y'^2)
+\frac{2}{3\sqrt5}x^3
\right)
\nonumber\\
&{}
+g^{-1}\left(x^2(\dot y^2-y'^2)
-\frac35x^2(\dot y^2+y'^2)\right)\,.
\end{align}
Since the $y$ excitations are massless we consider the case $p_{1-}=p_{2+}=0$. This has the effect that terms involving $\dot y^2-y'^2$ cannot contribute since they are proportional to $p_{1+}p_{2+}+p_{1-}p_{2-}$. In particular there is no contribution from exchange of $\varphi$. The contact diagram, s-channel and t,u-channel diagrams are easily computed and one finds
\begin{equation}
\mathcal A_c=-\frac65g^{-1}p_{1+}p_{2-}\,,\qquad
\mathcal A_s=-\frac45g^{-1}\frac{p_{1+}p_{2-}}{p_{1+}p_{2-}-1}\,,\qquad
\mathcal A_{t+u}=-\frac25g^{-1}p_{1+}p_{2-}\,.
\end{equation}
The total $yy\rightarrow xx$ amplitude is non-vanishing as claimed.

\section{Symmetric spaces with NSNS flux}\label{app:sym-NSNS}
In \cite{Wulff:2017hzy,Wulff:2017vhv} the absence of particle production in certain massless GKP scattering amplitudes was used as a criterion for integrability. Since the scattering of massless particles does not in general factorize even in an integrable theory one can question using such a condition to rule out integrability \cite{Hoare:2018jim}. There are various arguments that this condition nevertheless gives the right answer in the cases in question, e.g. the amplitudes considered do not have IR divergences (unlike the problematic ones in \cite{Hoare:2018jim}) and are well defined without any need to regulate them and the conditions one finds from absence of particle production reproduce what one finds by making an ansatz for the Lax connection \cite{Wulff:2015mwa}. It is nevertheless desirable to reproduce these conditions from scattering amplitudes involving massive particles were factorization is strictly necessary for integrability. Here we will go some way towards doing this by computing a certain (IR finite) 2-to-4 particle amplitude for two massless particles into two massless and two massive particles. Since two massless particles with opposite momenta can be separated asymptotically from each other and from the massive particles in the final state the usual arguments \cite{Zamolodchikov:1978xm,Parke:1980ki} suggest, at least at a heuristic level, that the existence of higher conserved charges should force such particle production amplitudes to vanish, at least when they are free of ambiguities related to IR divergences.

The backgrounds in question are the symmetric space solutions of type II supergravity \cite{Wulff:2017zbl} with non-zero B-field on the compact directions of the form
\begin{equation}
B=R^{-1}x^9(f_1e^3\wedge e^4+f_2e^5\wedge e^6+f_3e^7\wedge e^8)\,,
\label{eq:B}
\end{equation}
with $R$ the AdS radius, $e^i=dx^i+\mathcal O(x^3)$ and the parameters $f_i$ restricted as\footnote{In the special case of $AdS_3\times S^2\times S^2\times T^3$ the $B$-field can take a more general form \cite{Wulff:2017hzy} but the analysis is essentially the same.}
\begin{equation}
\begin{array}{ll}
AdS_5\times S^2\times S^2\times S^1 & f_1=0\\
AdS_3\times\mathbbm{CP}^2\times S^2(H^2)\times S^1 & f_1=f_2\\
AdS_3\times\mathbbm{CP}^2\times T^3 & f_1=f_2=f_3\\
AdS_3\times S^2\times S^2\times S^2(H^2)\times S^1 & \mbox{No restriction}\\
\end{array}
\end{equation}

Let us recall the argument of \cite{Wulff:2017hzy,Wulff:2017vhv}. Looking at the amplitude for $x_5x_6\rightarrow x_7x_7x_9$, where $x_{5,6}$ and $x_{7,8}$ belong to different factors in the geometry, one sees from the Lagrangian (\ref{eq:L3}) and (\ref{eq:L4}) that, in the low-energy limit, the only relevant interaction terms are the ones coming from the $B$-field, so that the only contribution to the amplitude comes from the Feynman diagram
$$
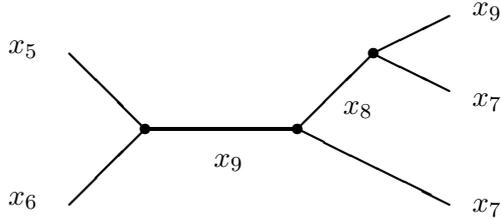
\setlength{\unitlength}{1cm}
\begin{picture}(17,5)(-1.5,0)
\thicklines
\put(3.2,3){$x_5$}
\put(3.2,1){$x_6$}
\put(4,3){\line(1,-1){1}}
\put(4,1){\line(1,1){1}}
\put(5,2){\circle*{0.15}}
\put(5,2){\line(1,0){2}}
\put(5.9,1.5){$x_9$}
\put(7,2){\circle*{0.15}}
\put(7,2){\line(1,1){1}}
\put(7.6,2.2){$x_8$}
\put(8,3){\line(2,1){1}}
\put(8,3){\line(2,-1){1}}
\put(8,3){\circle*{0.15}}
\put(9.3,3.5){$x_9$}
\put(9.3,2.3){$x_7$}
\put(7,2){\line(2,-1){2}}
\put(9.3,0.9){$x_7$}
\end{picture}
$$
One finds the condition $f_2f_3=0$ for this amplitude to vanish and by symmetry $f_if_j=0$ for any $i,j=1,2,3$ such that $f_i$ and $f_j$ are associated with flux on \emph{different} irreducible factors in the geometry.

Let us see if we can reproduce the same condition by looking instead at the closely related amplitude for $x_5x_6\rightarrow x_7x_8\varphi$, with four massless and one massive particle. There are five diagrams that contribute
$$
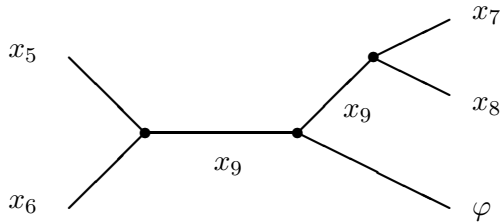
\setlength{\unitlength}{1cm}
\begin{picture}(17,5)(-1.5,0)
\thicklines
\put(3.2,3){$x_5$}
\put(3.2,1){$x_6$}
\put(4,3){\line(1,-1){1}}
\put(4,1){\line(1,1){1}}
\put(5,2){\circle*{0.15}}
\put(5,2){\line(1,0){2}}
\put(5.9,1.5){$x_9$}
\put(7,2){\circle*{0.15}}
\put(7,2){\line(1,1){1}}
\put(7.6,2.2){$x_9$}
\put(8,3){\line(2,1){1}}
\put(8,3){\line(2,-1){1}}
\put(8,3){\circle*{0.15}}
\put(9.3,3.5){$x_7$}
\put(9.3,2.3){$x_8$}
\put(7,2){\line(2,-1){2}}
\put(9.3,0.9){$\varphi$}
\end{picture}
$$
or a similar diagram with the $\varphi$-line attached to one of the four massless external legs instead. We consider the kinematic region $p_{1-}=p_{2+}=p_{3-}=p_{4+}=0$ so that the massless particles in the in- and out-state travel in opposite directions and are separated asymptotically. Energy-momentum conservation implies that the momentum for the massive particle satisfies $p_{5+}=p_{1+}-p_{3+}$ and $p_{5-}=p_{2-}-p_{4-}$ while the mass-shell condition for $\varphi$ is $p_{5+}p_{5-}=1$. From (\ref{eq:B}) we see that the $x_5x_6x_9$-vertex is proportional to $p_+p'_--p'_+p_-$ where $p$($p'$) is the $x_5$($x_6$) momentum and similarly for the $x_7x_8x_9$-vertex. Finally the $x_9x_9\varphi$-vertex in (\ref{eq:L3}) is proportional to $p_+p_+'+p_-p_-'$ where $p$ and $p'$ are the two $x_9$ momenta. Using these facts one finds that the amplitude is proportional to (each term corresponding to one diagram)
\begin{align}
&
\frac{p_{3+}p_{4-}[p_{1+}(p_{1+}-p_{5+})+p_{2-}(p_{2-}-p_{5-})]}{(p_{1+}-p_{5+})(p_{2-}-p_{5-})}
-\frac{p_{1+}p_{2-}p_{3+}p_{4-}}{p_{5-}(p_{2-}-p_{5-})}
\nonumber\\
&{}
-\frac{p_{1+}p_{2-}p_{3+}p_{4-}}{p_{5+}(p_{1+}-p_{5+})}
+\frac{p_{3+}p_{4-}(p_{3+}+p_{5+})}{p_{5-}}
+\frac{p_{3+}p_{4-}(p_{4-}+p_{5-})}{p_{5+}}\,,
\end{align}
which vanishes by momentum conservation alone, without taking the massive particle ($\varphi$) momentum $p_5$ on-shell. Since the other massive modes ($z$) don't couple to $x$ directly but only through $\varphi$ via the $zz\varphi$-vertex this implies that the amplitude for $x_5x_6\rightarrow x_7x_8zz$ also vanishes.

The next simplest amplitude to consider is $x_5x_6\rightarrow x_7x_8\varphi\varphi$. We will now show that the vanishing of this, and similar, amplitudes leads to the same conditions derived in \cite{Wulff:2017hzy,Wulff:2017vhv} from vanishing of the massless particle production amplitudes discussed above.

First we note that, for the same reason that the amplitude for $x_5x_6\rightarrow x_7x_8zz$ vanishes, the contributions involving the $\varphi^3$-vertex vanish. The contributing diagrams therefore come from taking the basic $x_5x_6\rightarrow x_7x_8$ diagram
$$
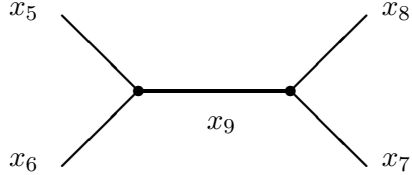
\setlength{\unitlength}{1cm}
\begin{picture}(9,4)(1,0)
\thicklines
\put(3.3,3){$x_5$}
\put(3.3,1){$x_6$}
\put(4,3){\line(1,-1){1}}
\put(4,1){\line(1,1){1}}
\put(5,2){\circle*{0.15}}
\put(5,2){\line(1,0){2}}
\put(5.9,1.5){$x_9$}
\put(7,2){\circle*{0.15}}
\put(7,2){\line(1,1){1}}
\put(7,2){\line(1,-1){1}}
\put(8.2,3){$x_8$}
\put(8.2,1){$x_7$}
\end{picture}
$$
and inserting either a $\varphi xx$-vertex on two of the lines or a $\varphi\varphi xx$-vertex on one line. There are five diagrams of the latter type and $15$ of the former, 10 with insertions on different lines and 5 with two insertions on the same line, for a total of 20 diagrams. A direct calculation shows that the amplitude is non-vanishing. Specifically, denoting the contribution of a diagram with insertions on lines $i$ and $j$ as $[ij]$, with $i=1,2,3,4$ for the external lines and $i=0$ for the internal line, the contributions coming from  diagrams with insertions on the same line are
\begin{align}
{}[00]=&{}\,\frac{(p_{1+}(p_{1+}-p_{5+})+p_{2-}(p_{2-}-p_{5-}))[(p_{1+}-p_{5+})p_{3+}+(p_{2-}-p_{5-})p_{4-}]}{(p_{1+}-p_{5+})(p_{2-}-p_{5-})}\,,\nonumber\\
{}[11]=&{}\,\frac{p_{1+}p_{2-}[(p_{1+}-p_{5+})p_{3+}+p_{5-}(p_{5-}+p_{6-})]}{(-p_{5-})(-p_{5-}-p_{6-})}\,,\nonumber\\
{}[22]=&{}\,\frac{p_{1+}p_{2-}[(p_{2-}-p_{5-})p_{4-}+p_{5+}(p_{5+}+p_{6+})]}{(-p_{5+})(-p_{5+}-p_{6+})}\,,\nonumber\\
{}[33]=&{}\,\frac{p_{3+}p_{4-}[p_{1+}(p_{3+}+p_{6+})+(p_{5-}+p_{6-})p_{6-}]}{(p_{5-}+p_{6-})p_{6-}}\,,\nonumber\\
{}[44]=&{}\,\frac{p_{3+}p_{4-}[p_{2-}(p_{4-}+p_{6-})+(p_{5+}+p_{6+})p_{6+}]}{(p_{5+}+p_{6+})p_{6+}}\,.
\end{align}
This is up to an overall constant coefficient which we have not kept track of and the final expression should be symmetrized in $p_5$ and $p_6$. From diagrams with insertions on different lines we get
\begin{align}
{}[03]=&{}\,\frac{p_{3+}p_{4-}[p_{1+}(p_{3+}+p_{6+})+p_{2-}(p_{4-}+p_{6-})]}{(p_{4-}+p_{6-})p_{6-}}\,,\nonumber\\
{}[04]=&{}\,\frac{p_{3+}p_{4-}[p_{1+}(p_{3+}+p_{6+})+p_{2-}(p_{4-}+p_{6-})]}{(p_{3+}+p_{6+})p_{6+}}\,,\nonumber\\
{}[12]=&{}\,\frac{p_{1+}p_{2-}(p_{3+}+p_{6+})(p_{4-}+p_{5-})}{p_{5-}p_{6+}}\,,\nonumber\\
{}[10]=&{}-\frac{p_{1+}p_{2-}[(p_{3+}+p_{6+})p_{3+}+(p_{4-}+p_{6-})p_{4-}]}{p_{5-}(p_{4-}+p_{6-})}\,,\nonumber\\
{}[13]=&{}-\frac{p_{1+}p_{2-}p_{3+}p_{4-}(p_{3+}+p_{6+})}{p_{5-}p_{6-}(p_{4-}+p_{6-})}\,,\nonumber\\
{}[14]=&{}-\frac{p_{1+}p_{2-}p_{3+}p_{4-}}{p_{5-}p_{6+}}\,,\nonumber\\
{}[20]=&{}-\frac{p_{1+}p_{2-}[(p_{3+}+p_{6+})p_{3+}+(p_{4-}+p_{6-})p_{4-}]}{p_{5+}(p_{3+}+p_{6+})}\,,\nonumber\\
{}[23]=&{}-\frac{p_{1+}p_{2-}p_{3+}p_{4-}}{p_{5+}p_{6-}}\,,\nonumber\\
{}[24]=&{}-\frac{p_{1+}p_{2-}p_{3+}p_{4-}(p_{4-}+p_{6-})}{p_{5+}p_{6+}(p_{3+}+p_{6+})}\,,\nonumber\\
{}[34]=&{}\,\frac{p_{3+}p_{4-}(p_{3+}+p_{5+})(p_{4-}+p_{6-})}{p_{5-}p_{6+}}\,.
\end{align}
Adding them up and symmetrizing in $(p_5,p_6)$ there are many cancellations and we find the total contribution from cubic interaction terms
\begin{equation}
\mathcal A_3\propto2p_{1+}p_{4-}+2p_{2-}p_{3+}+6p_{1+}p_{2-}+6p_{3+}p_{4-}\,.
\end{equation}
The contribution from diagrams with a quartic vertex is
\begin{align}
\mathcal A_4\propto
p_{1+}p_{4-}+p_{2-}p_{3+}
+2p_{1+}p_{2-}
+2p_{3+}p_{4-}\,.
\end{align}
It is easy to see that the two cannot cancel regardless of the (different) proportionality factors in front. The only way the amplitude can vanish is therefore if the overall proportionality factor vanishes. This factor is proportional to $f_2f_3$, just as for the purely massless amplitude discussed above, so we get the same condition, namely $f_2f_3=0$. Similarly we get all the other conditions one obtains by requiring the massless 2-to-3 amplitude to vanish, as claimed.

\bibliographystyle{nb}
\bibliography{biblio}{}

\end{document}